\documentclass[aps,prl,twocolumn,superscriptaddress,showpacs,floatfix]{revtex4}
\usepackage{xspace}

\newcommand{\dd}{\ensuremath{\text{d}}} 

\newcommand{\Eu}{\ensuremath{\text{e}}}
\newcommand{\kb}{\ensuremath{k_\text{B}}}

\newcommand{\Tc}{\ensuremath{T_\text{c}}\xspace}

\newcommand{\Eq}[1]{Eq.~(\ref{eqn:#1})}
\newcommand{\Fig}[1]{Fig.~\ref{fig:#1}} 

\newcommand{\Na}[1]{\ensuremath{^{#1}}\text{Na}}

\newcommand{\vct}[1]{\ensuremath{\boldsymbol{#1}}}

\newcommand{\nex}{\ensuremath{n_\text{ex}}\xspace}
\newcommand{\nc}{\ensuremath{n_\text{c}}\xspace}

\newcommand{\Nex}{\ensuremath{N_\text{ex}}\xspace}

\newcommand{\cf}{\textit{cf.~}}

\usepackage{graphicx}
\usepackage{amsmath}
\usepackage{amssymb}
\makeatletter

\newcommand{\Rmnum}[1]{\expandafter\@slowromancap\romannumeral #1@}
\makeatother
\begin{document}
\title{Sound propagation in a Bose-Einstein condensate at finite temperatures}
\author{R.~Meppelink, S.~B.~Koller, and P.~van der Straten}
\affiliation{Atom Optics and Ultrafast Dynamics, Utrecht University,\\ P.O. Box 80,000, 3508 TA Utrecht, The Netherlands}
\date{\today}
\begin{abstract}
  We study the propagation of a density wave in a magnetically trapped Bose-Einstein condensate at finite temperatures. The thermal cloud is in the hydrodynamic regime and the system is therefore described by the two-fluid model. A phase-contrast imaging technique is used to image the cloud of atoms and allows us to observe small density excitations. The propagation of the density wave in the condensate is used to determine the speed of sound as a function of the temperature. We find the speed of sound to be in good agreement with calculations based on the Landau two-fluid model.
\end{abstract}
\maketitle
\section{Introduction}
Long wavelength excitations of a Bose-Einstein condensate (BEC) with repulsive interactions exhibit a phononlike, linear dispersion, causing these excitations to move at a finite speed $c$, the speed of sound. Excitations with a wavelength comparable to the size of the cloud result in collective shape oscillations of the system \cite{PhysRevLett.81.500}. 
%In the previous chapter excitation with wavelengths comparable to the axial size of the BEC are studied and the time-dependent behavior is characterized by shape-dependent oscillatory modes. %These excitations do not directly probe the phononic nature of the excitations since the harmonic confinement causes discretization of the eigen frequencies. 
%CHAPTER The dispersion relation is linear up to wave vectors of the order of the inverse of the healing length $\xi = 1/\sqrt{8 \pi a n}\approx 0.1 $ \micron, where $a$ is the scattering length and $n$ is the condensate density. Perturbations with a wavelength larger than several micrometer, but much shorter than the axial size of the cloud, should therefore give rise to the excitation of phononlike quasiparticles, propagating at the speed of sound.
The dispersion relation is linear up to wave vectors of the order of the inverse of the healing length. Perturbations with a wavelength much shorter than the axial size of the cloud and larger than the healing length therefore give rise to the excitation of a hydrodynamic mode propagating at the speed of sound.

In the case of liquid helium sound excitations have been extensively studied below the $\lambda$-point, where a superfluid and a normal fluid coexist. In this regime two sound modes can be distinguished: the first mode, first sound, consists of an in-phase oscillation of the superfluid and normal fluid component, while the second mode, second sound, consists of an out-of-phase oscillation of the superfluid and normal fluid component. The occurrence of two distinct modes is caused by the presence of both a superfluid density and normal fluid density, which are coupled.  One of the drawbacks of liquid helium is that the interactions are so strong that a clear distinction between the two components is difficult which complicates the interpretation of the phenomena.

In the dilute gaseous BEC studied here a two-fluid system exists below \Tc, analogous to the case of liquid helium. In our setup the thermal cloud is in the hydrodynamic regime, where collisions between atoms are rapid enough to establish a state of dynamic local equilibrium in the non-condensed atoms. In contrast to Bose-condensed liquid helium, the superfluid in the gaseous BEC corresponds directly to the Bose-condensed atoms and the normal component directly to the thermal, non-condensed atoms, since the interactions are much weaker.

First and second sound in a Bose gas exhibit different features than those in a Bose liquid \cite{proc}. In liquid helium, the coupling between the density and temperature is weak, since $C_p\simeq C_v$, with $C_{p,v}$ the specific heat for constant pressure and constant volume, respectively. As a result, in liquid helium first sound is mainly a density wave, while second sound is an almost pure temperature wave. In contrast, in a Bose gas, the density and temperature fluctuations are strongly coupled, since $C_p/C_v \not\simeq 1$. The first sound mode in a Bose gas is largely an oscillation of the density of the thermal cloud (the normal fluid) and second sound is largely an oscillation of the density of the condensate (the superfluid) \cite{PhysRevA.58.4044}.  In superfluid helium, only the first sound mode can be excited by a density perturbation. In contrast, second sound has a significant weight in the density response function in superfluid Bose gases at finite temperatures and can be excited by a local perturbation of the density. 

Since we can directly image BECs and make a clear distinction between both components, it allows for a direct comparison with theoretical descriptions of the two-fluid system modeling the interactions between both components. Thus the research using weakly interacting Bose gases promises results that will go beyond the results obtained using liquid helium. 

In a pioneering paper by Lee and Yang the speed of first and second sound is derived for a dilute Bose gas, although there is no coupling between both sound modes \cite{PhysRev.113.1406}.  This interaction is taken into account in the two-fluid model developed by Landau for liquid helium \cite{JPhys.V.71} and in a hydrodynamic model developed by Zaremba, Griffin, and Nikuni for trapped Bose gases \cite{PhysRevA.57.4695}. In these papers it is shown that the hydrodynamic second sound mode at finite temperature extrapolates to the $T=0$ Bogoliubov phonon mode. 
%Furthermore, for low temperatures the velocities of first and second sound are close to each other. Both sound modes couple in the descriptions which incorporate interaction between the sound modes and as a consequence give rise to an avoided crossing.

The propagation of sound in a harmonically trapped, almost pure BEC in the collisionless regime has been observed experimentally in a pioneering experiment by the MIT group \cite{PhysRevLett.79.553,PhysRevLett.80.2967} and studied theoretically by various authors \cite{PhysRevA.57.518,PhysRevA.58.1563,PhysRevA.58.2385,PhysRev.113.1406,JPhys.V.71,JLowTempPhys.116.277}. After the first experiment, sound propagation has been observed for a BEC in an optical lattice, the excitation spectrum of a BEC has been measured and the excitation of shock waves is observed \cite{PhysRevA.70.023609,PhysRevLett.88.120407, PhysRevLett.101.170404}. 

The work presented here describes the experimental observation of a propagating sound wave in an elongated BEC at finite temperatures and extends the study by the MIT group in two ways. First, in the work presented here the propagation of a sound wave is observed at finite temperatures. The thermal cloud is in the hydrodynamic regime above $T_\text{c}$ and the cloud is therefore a two-fluid system below $T_\text{c}$: a superfluid BEC coexists with a normal fluid of thermal atoms. If the thermal cloud is already close to the hydrodynamic regime above $T_\text{c}$, it will be deeply in the hydrodynamic regime when the BEC forms, since the collision rate $\gamma$ is dominated by the collisions between the condensed and the thermal atoms determined by the rate $\gamma_{12}$ \cite{PhysRevA.65.011601}. In the two-fluid system interactions between the superfluid and the normal component are expected to play an important role. Second, the high signal-to-noise ratio of our imaging technique allows us to make smaller excitations than is used in the MIT experiments, thereby limiting non-linear effects. The large atom number BECs in combination with the weak axial confinement results in typical axial BEC lengths of more than 2 mm. This allows for the determination of the speed of sound with a high accuracy, since the propagation distance can be large before the sound wave reaches the edge of the BEC.

\section{Sound propagation in a dilute Bose-condensed gas}

Since second sound at finite temperatures extrapolates to the $T=0$ Bogoliubov phonon mode, we start this discussion in the $T=0$ limit. 
The speed of second sound in the absence of a thermal cloud can be derived using the Gross-Pitaevskii equation (GPE). Reformulated as a pair of hydrodynamic equations, neglecting the quantum pressure and after linearization, the GPE can be written in the simplified, hydrodynamic form \cite{PhysRevA.58.2385}
\begin{equation}
  \frac{ \partial^2 \delta n}{\partial t^2} = \nabla  \left( c^2(\vec{r}) \nabla \delta n \right),
  \label{hydrolin}
\end{equation}
where the departure of the density from its equilibrium density $n_\text{eq}$ is given by $\delta n(\vec{r},t)  = n(\vec{r},t) - n_\text{eq}(\vec{r})$ and the local speed of sound $c(\vec{r})$ is defined by
\begin{equation}
  m c^2(\vec{r}) = \mu -V_\text{ext}(\vec{r}),
  \label{eqn:muismc2}
\end{equation}
where $\mu$ is the chemical potential, $V_\text{ext}$ is the external confinement and $m$ is the mass. In a uniform Bose gas, $V_\text{ext}=0$, the speed of second sound is given by $c=\sqrt{\mu / m}=\sqrt{g\,n_\text{c}/m}$, with \nc the condensate density. This result was first derived by Lee and Yang \cite{PhysRev.113.1406} based on theory developed by Bogoliubov \cite{JPhysUSSR.11.23} and is therefore often referred to as the Bogoliubov speed of sound, which we will refer to as $c_\text{B}$ in this paper. $c_\text{B}$ only depends on temperature through the BEC density and is independent of the thermal density. 

The experiments are conducted in an elongated 3D trap where the external confinement in the radial (subscript rad) and axial (subscript ax) direction is given by 
\begin{equation}
  V_\text{ext}(x,y,z)=\frac{1}{2} m \left(\omega^2_\text{rad} x^2 + \omega^2_\text{rad} y^2 + \omega^2_\text{ax} z^2   \right),
  \label{eqn:vext}
\end{equation}
with $\omega_\text{rad} \gg \omega_{ax}$, and the density depends on the position. Although the confinement is highly anisotropic, the BEC is not fully in the 1D regime, since $\mu \gg \hbar \omega_\text{rad}$. 
For the radial average density we use the Thomas-Fermi (TF) value $\bar{n}(z) = n(0,0,z)/2$, where $n(0,0,z)$ is the central axial density.
The local Bogoliubov speed of sound is given in terms of the radial average density by
\begin{equation}
  c_\text{B}(z) =  \sqrt{\frac{g\,\bar{n}(z)}{m}} = \sqrt{\frac{g\,n(0,0,z)}{2 m}},
  \label{eqn:bogolsostrap}
\end{equation}
a result confirmed using different methods \cite{PhysRevLett.80.2967,PhysRevA.57.518,PhysRevA.58.1563,PhysRevA.58.2385}. In our experiment the axial confinement is weak and the density varies slowly along the axial direction.

To our knowledge two theoretical descriptions are available in which the effect of the interaction between the normal and the superfluid component on the speed of sound is taken into account. Zaremba, Griffin, and Nikuni have derived the two-fluid hydrodynamic equations for weakly interacting Bose gases \cite{PhysRevA.57.4695} and use them to discuss first and second sound for a uniform Bose gas \cite{PhysRevA.56.4839}. We refer to this description as the ZGN model. In the same paper, they derive the Landau two-fluid equations for a dilute gas in a complete local equilibrium and use these equations to calculate the first and second sound velocities. We refer to this description as the Landau model. In Ref.~\cite{JLowTempPhys.116.277} it is shown the ZGN model is valid in the limit in which collisions between the condensed and the non-condensed atoms are ignored on the timescale of the collective excitation $\gamma_\mu/\omega \rightarrow 0$. The Landau model developed for dense fluids such as superfluid helium is valid for dilute Bose gases in the opposite limit of complete local equilibrium $\gamma_\mu/\omega \rightarrow \infty$ \cite{JLowTempPhys.116.277}. Here, $\gamma_\mu$ is the relaxation rate for chemical potential differences between the condensate and the thermal cloud as given in Ref.~\cite{JLowTempPhys.116.277} and $\omega$ is the excitation frequency. 

We introduce a measure for the hydrodynamicity of the thermal cloud in the axial direction $\bar{\gamma}\equiv \gamma_{22}/\omega_\text{ax}$, where the collision rate $\gamma_{22} = n_\text{eff} \,\sigma\, v_\text{rel}$ is the average number of collisions in the thermal cloud. Here, the relative velocity $v_\text{rel}=\sqrt{2} \bar{v}_\text{ex}$, where $\bar{v}_\text{ex}=\sqrt{8\kb T/m \pi}$ is the thermal velocity at temperature $T$ and $m$ is the mass and $\sigma =8\pi a^{2}$ is the isotropic cross-section of two bosons with $s$-wave scattering length $a$. Furthermore, $n_\text{eff}=\int n_\text{ex}^2\left (\vct{r}\right )\dd V/\int n\left(\vct{r}\right )\dd V = n0_\text{ex}/\sqrt{8}$ for an equilibrium distribution in a harmonic potential, where $n0_\text{ex}$ is the peak density. Written in terms of the number of thermal atoms  $\Nex=n0_\text{ex} \left ( 2 \pi \kb T/ \left ( m \bar{\omega}^2  \right ) \right )^{3/2}$ and the geometric mean of the angular trap frequencies $\bar{\omega}^3 \equiv \omega^2_\text{rad} \omega_\text{ax}$, this results in $\gamma_{22}= \Nex m \sigma \bar{\omega}^3/(2 \pi^2 \kb T) \approx~{90}{{\rm s}^{-1}}$ for the highest number of atoms and corresponds to a hydrodynamicity of  $\bar{\gamma}\lesssim~{10}$ in the axial direction. We have observed the transition from the collisionless regime to the hydrodynamic regime by studying a thermal dipole mode above \Tc \cite{onzePRL}. Furthermore, it is noted in Ref.~\cite{PhysRevA.65.011601} that the thermal cloud is deeper in the hydrodynamic regime when a BEC forms due to collisions with condensed atoms. As a result, the thermal cloud is even more hydrodynamic below \Tc than it is above the transition temperature.

Both models are used to calculate the speed of first and second sound as solutions of an equation of the form $u^4-A u^2 +B =0$, where $A$ and $B$ are coefficients which are given for both models in Ref.~\cite{JLowTempPhys.116.277}. These coefficients depend on the condensate density \nc, the non-condensate density \nex and the temperature $T$. In Ref.~\cite{JLowTempPhys.116.277} the differences between both models for constant density $n=\nc+\nex$ are found to be very small in the case of a weakly interacting Bose gas. As a result, the transfer of atoms required to equilibrate the condensed and non-condensed atoms is found to play a minor role in the determination of the speed of first and second sound \cite{PhysRevA.56.4839}.

\begin{figure}[htpb]
  \begin{center}
     \includegraphics[width=0.45\textwidth]{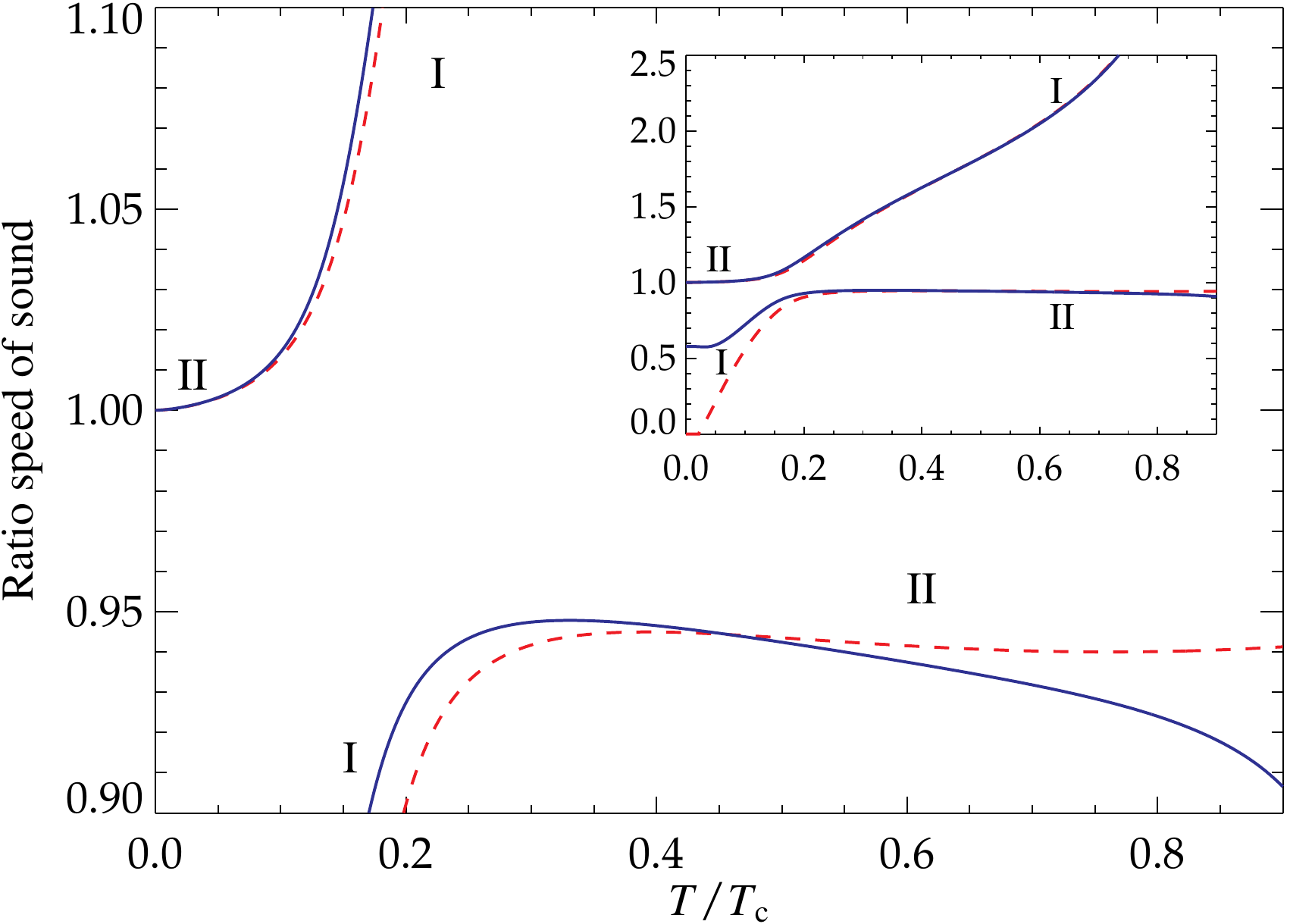}
  \end{center}
  \caption{(Color online) The speed of first (\Rmnum{1}) and second (\Rmnum{2}) sound calculated in the Landau model (solid lines) and the ZGN model (dashed lines), normalized to $c_\text{B}(0)$ (\Eq{bogolsostrap}) as a function of the reduced temperature $T/T_\text{c}$. The inset shows the same temperature range on a larger vertical scale. The densities and temperature used for the calculations for $0.2<T/\Tc<0.9$ correspond to the intrapolated experimental values. For $T/\Tc<0.2$ \nc is kept constant, and \nex is extrapolated to $\nex=0$ for $T=0$.}
  \label{fig:ZGNtheory}
\end{figure}

The speed of first and second sound for both models is shown in \Fig{ZGNtheory}, where close to \Tc the first (second) sound mode mainly corresponds to the density wave in the thermal (condensed) fraction. The figure shows the coupling between both sound modes cause the speed of second sound to be smaller than the Bogoliubov speed of sound. Around $T=0.15\Tc$ this coupling results in an avoided crossing. Note that the position of the avoided crossing in \Fig{ZGNtheory} is outside the region of validity of the model, which is valid for $\kb T \gg \mu$ \cite{JLowTempPhys.116.277}. 

\section{Experiment}
The experimental setup used to create large number BECs is described in Ref.~\cite{RevSciInstr.78.013102}. In short, \Na{23} atoms are cooled and trapped in a dark-spot MOT and transferred to a magnetic trap (MT) after being spin-polarized \cite{PhysRevA.73.063412}. Forced evaporative cooling on these atoms yields roughly $1\cdot10^9$ thermal atoms around $T=T_\text{c}$. In order to prevent three-body losses, which limit the density, as well as to increase the collision rate with respect to the axial trap frequency, we work with axially decompressed traps and have reached the hydrodynamic regime in the thermal cloud \cite{PhysRevA.75.031602}. The resulting elongated, cigar-shaped clouds used for the experiments described here have an aspect ratio $\omega_\text{rad}/\omega_\text{ax} \approx 65$. 

The experiments are conducted on clouds at various temperatures below $T_\text{c}$. The number of condensed atoms, slightly depending on the temperature, is roughly $N_\text{c} = 1.7 \cdot 10^8$ and the BEC density is about $2.5 \cdot 10^{20} \text{m}^{-3}$. At the lowest temperatures, the  BEC has a radial TF radius of roughly 22 $\mu$m  and an axial TF radius of 1.4 mm.

The clouds are imaged using phase-contrast imaging, the details of which are described in Ref \cite{PhysRevA.pci}. Briefly, the atoms are imaged \textit{in situ} and in contrast to other implementations of the phase-contrast imaging technique \cite{Science.273.84}, our implementation does not aim at non-destructive imaging, but uses the periodicity of the intensity of the phase contrast imaging technique as a function of the accumulated phase of a probe beam that propagates through a cloud of atoms. Therefore, the intensity signal for large enough accumulated phase show rings in the intensity profile. The number of rings is a sensitive measure for the atomic density and allows us to determine the BEC density, the thermal density and the temperature within a few percent. The lens setup used results in a diffraction limited resolution of roughly 4 $\mu$m.

\begin{figure}[htpb]
  \begin{center}
     \includegraphics[width=0.25\textwidth]{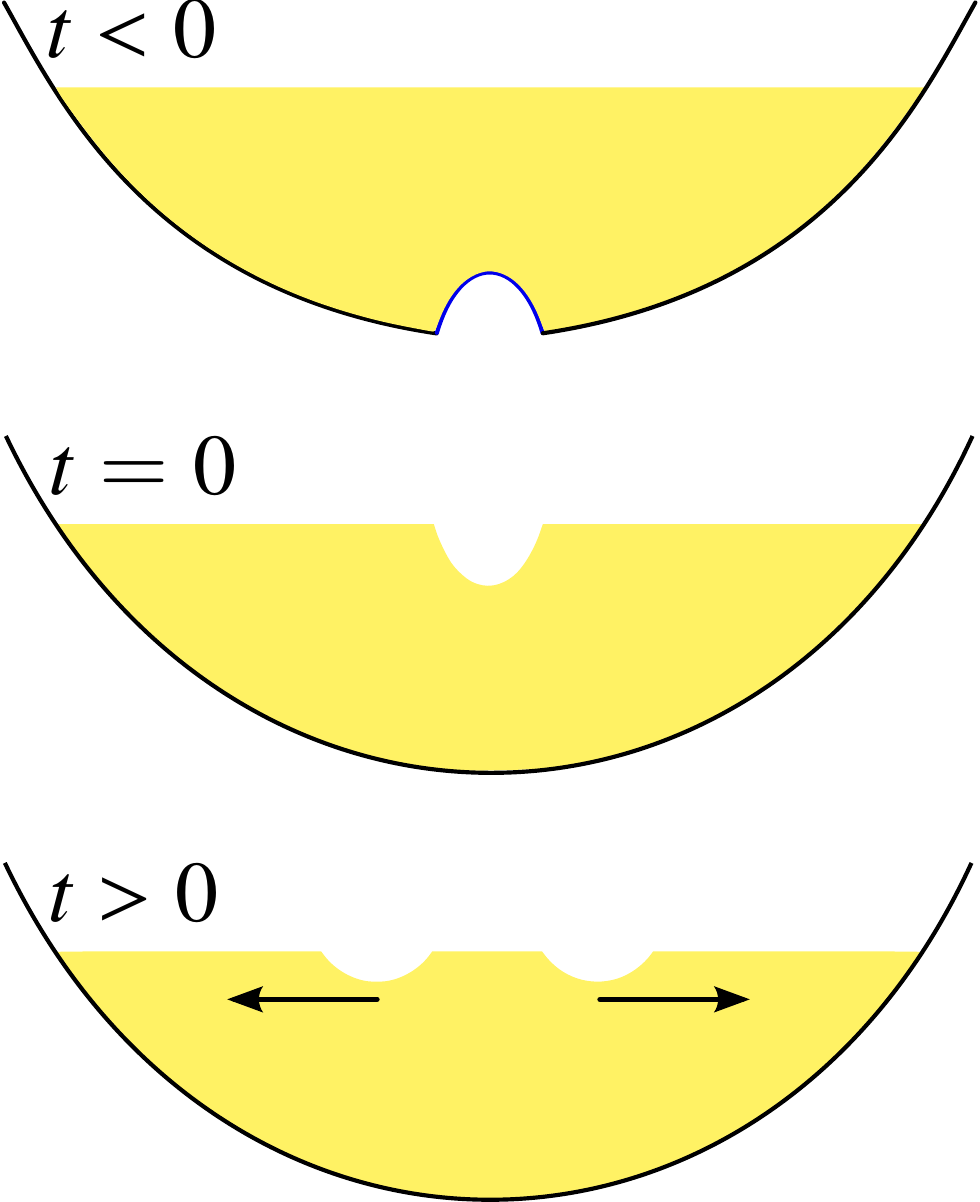}
  \end{center}
  \caption{(Color online) Schematic representation of the excitation of a sound wave, where the trapping potential, height and width of the perturbation are roughly on scale. At $t<0$ a BEC is formed in the presence of a blue-detuned repulsive dipole beam which adjusts the trapping potential. At $t=0$ the dipole beam is suddenly turned off and the inflicted density perturbation causes two density dips to move outward for $t>0$, propagating at the speed of sound.}
  \label{fig:sndpropschematic}
\end{figure}

For the experiments we evaporatively cool the atoms in the presence of a blue-detuned focused laser beam aimed at the center of the trap, which acts as a repulsive optical dipole trap. The beam is focused using a cylindrical lens yielding a sheet of light with a ($1/\Eu$)-width of the intensity of $127 \pm 20 \, \mu$m. The non-focused direction has a ($1/\Eu$)-width of roughly 5 mm. The light of the dipole beam is detuned 20 nm below the \Na{23} D$_2$ transition and the power adjusted in such a way that the repulsive potential has a height of $\left ( 0.24 \pm 0.04 \right ) \mu$, where $\mu$ is the chemical potential. Due to the large detuning heating caused by light scattering is negligible. The laser power of the dipole beam is continuously measured and used in an electronic feedback circuit controlling the efficiency of the acousto-optical modulator (AOM) that deflects laser light into an optical fiber whose output is used for the dipole beam. Using this procedure the stabilized intensity has a RMS fluctuation of 0.1\%. Typically, the power is adjusted on the order of milliseconds and the power is therefore not stabilized instantaneously, mainly due to heating of the optical fibers. However, the height of the potential is not critical during the first stages of the evaporation (tens of seconds), where the thermal energy $\kb T$ is large with respect to the dipole potential, ensuring the stability of the height of the potential long before the BEC is formed. An alternative way of exciting a sound wave is by turning on the repulsive potential after the BEC is formed. Since this procedure is immediately sensitive to the height of the potential this alternative results in less reproducible excitations than the procedure used in our experiments. Turning the dipole beam suddenly off ((1/\Eu)-time $\sim$ 250 ns) causes a local dip of the BEC density. This perturbation splits up in two waves propagating symmetrically outward, both with half the amplitude of the initial perturbation. A schematic representation of the excitation procedure is shown in \Fig{sndpropschematic}.

Clouds are imaged at about ten different times after the dipole beam is switched off, where for each shot a new cloud is prepared, since the imaging scheme used is destructive. The initial conditions of the newly prepared clouds show only a small variation, since the density in the final stage of the evaporative cooling process is limited by three-body losses. The density as a function of the axial position is determined by making 1D fits of the radial profile for all axial positions. The fit function used is a bimodal distribution which is the sum of a Maxwell-Bose distribution describing the thermal cloud and a TF distribution describing the BEC. Each propagation time for a series is mostly measured twice. The total accumulated phase as well as the radial width of the BEC can be used as a measure for the local density, although the width is expected to yield inferior results due to lensing effects and the limited resolution in the radial direction. 

\begin{figure}[htpb]
  \begin{center}
     \includegraphics[width=0.3\textwidth]{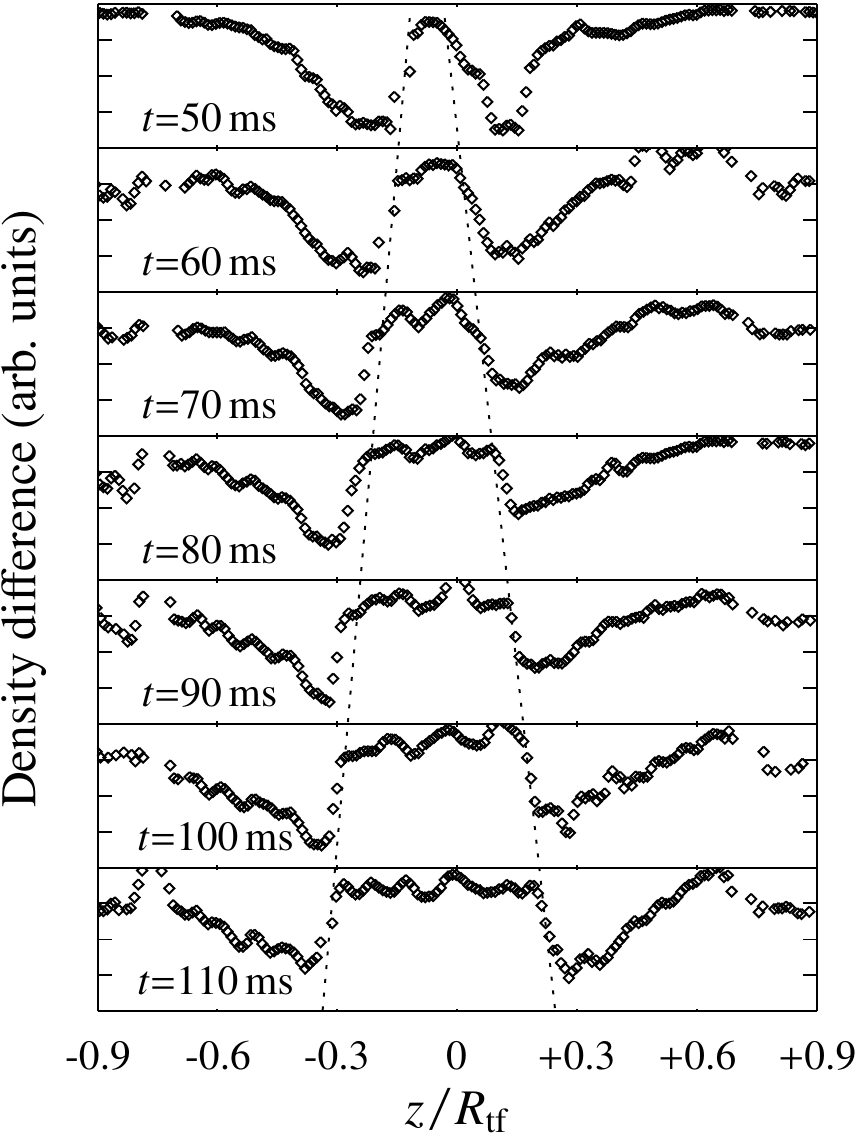}
  \end{center}
  \caption{Density profiles showing the peak density of the BEC as a function of the axial position for seven different propagation times. The unperturbed density profile is subtracted from  perturbed ones to increase the visibility of the density dips. The dotted line is a guide to the eye following the trailing edge of the dip. The horizontal scale is given in terms of the axial TF radius $R_\text{tf}$, where $R_\text{tf} \approx 1.4$ mm.}
  \label{fig:expdens}
\end{figure}

\begin{figure}[htpb]
  \begin{center}
     \includegraphics[width=0.49\textwidth]{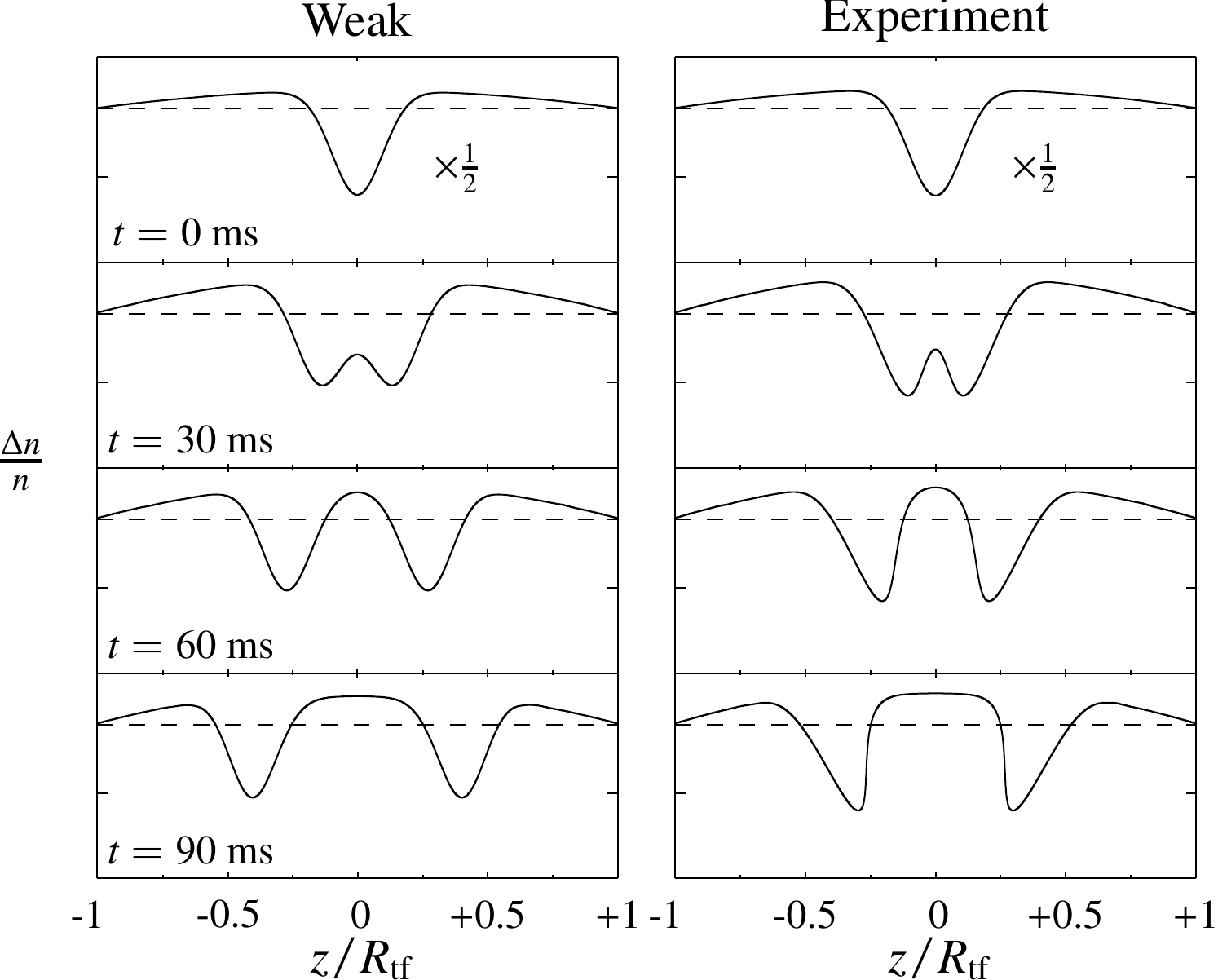}
  \end{center}
  \caption{Simulation of the normalized density profiles $\Delta n/n$ along the axial axis in terms of the axial TF radius $R_\text{tf}$ for different times $t$ after the sound wave is excited. The left column corresponds to a weak perturbation $\Delta n/n = 10^{-4}$, the right column corresponds to the perturbation applied in the experiment, $\Delta n/n = 0.25 $ and therefore both columns are on a different vertical scale. Note that the $t=0$ figures (top row) have a vertical scale twice as large as the other figures in the same column. The dashed lines in the figures indicate $\Delta n/n=0$.}
  \label{fig:simdens}
\end{figure}

\section{Results \& discussion}
The resulting axial density profiles of the condensate, shown in \Fig{expdens} for various propagation times, clearly shows the outward traveling density dips. In the density profiles of the thermal cloud these dips are absent, and no propagation is observed. The propagation of sound in the thermal cloud is discussed at the end of this section and for now we consider only the density perturbation of the condensed fraction of the cloud. 

The density profiles shown in \Fig{expdens} show that the shape of the perturbation, which is initially approximately Gaussian, deforms during the propagation. The deformation is caused by the dependence of the speed of sound on the density and its effect will be estimated using \Eq{bogolsostrap}. The estimated change of the sound velocity corresponding to a change in density $\Delta n$ is approximated by $\Delta c = (c/2) \Delta n/n$. The dependence of the sound velocity on the density  causes the center of the dip to move slower than the edges. Furthermore, this effects results in the trailing edge to 'overtake' the center, while the leading edge outruns the center part. The resulting high density gradient leads to the formation of shock waves when the density gradient is of the order of 1/$\xi$, where $\xi$ is the healing length \cite{PhysRevLett.101.170404}. Formation of shock waves complicates the propagation due to strong non-linear behavior. The typical time for the formation of shock waves is estimated by considering the difference in traveled distance $\Delta z$ of the tailing edge (density $n$) with respect to the center (density $n-\Delta n$) after $\tau$ propagation time, yielding $\Delta z \approx -(c \tau/2) \Delta n/n$ \cite{PhysRevA.58.1563}. The edge will reach the center when $\Delta z \sim \sigma$, where $\sigma$ is the width of the perturbation, resulting in $\tau \gtrsim \sigma n /(\Delta n \,c)$ which is the time for shock waves to form. For our typical parameters this yields $\tau \gtrsim 200$ ms. In the experiments propagation times are less than 110 ms.

For times less than $\tau$ we already see that the propagating wave deforms. We have made a simulation of the Gross-Pitaevskii (GPE) equation which describes the BEC at $T=0$ to analyze the effect the deformation has on the propagation of the condensate density wave. The GPE is numerically solved using the time-splitting spectral method described in Ref. \cite{JOCP.187.318}. Since the experiments are done on very elongated, cigar-shaped BECs and all effects are found in the axial directions, the calculation time can be reduced by solving an effective 1D equation in the limit of strong coupling \cite{PhysRevA.65.043614}. Many experiments on both the statics and dynamics of BECs have shown that experiments can be modeled accurately by numerically solving the GPE, for example in experiments on interferometry \cite{PhysRevA.73.013604} and superfluidity \cite{RevModPhys.71.463}. Note that we assume that the temperature dependence on the deformation of the condensate density wave can be neglected. 

The simulated density profiles are shown in \Fig{simdens}, where the perturbation is excited by growing the BEC in the presence of an additional Gaussian shaped potential. At $t=0$ this extra potential is suddenly switched off, but the harmonic confinement remains. Two situations are shown; a small initial perturbation and a perturbation corresponding to the situation in the experiments. The simulations show the deformation of the initial shape of the perturbation and the steepening of the trailing edge of the condensate density wave under the experimental conditions. After about 200 ms the simulations indeed show the formation of shock waves. 

We have run simulations for larger perturbations $\Delta n/n \approx 0.5$ and find shock waves to form within a few tens of milliseconds. Comparing these results to the sound propagation experiments by the MIT group \cite{PhysRevLett.79.553,PhysRevLett.80.2967}, where applied perturbations are reported as large as $\Delta n/n=1$, suggests shock waves have formed in their experiments shortly after the sound wave is excited. 
The MIT group has measured the condensate density wave to propagate at $c_\text{B}(0)$, even though strong non-linear effects are expected to influence the propagation, as already remarked by Kavoulakis \textit{et al.} \cite{PhysRevA.58.1563}.  

In our experiment sound waves are excited in elongated cigar-shaped BECs allowing us to make a perturbation much longer than the radial size to ensure the excitation of a one-dimensional motion, while it remains small with respect to the axial size. In the experiment reported by the MIT group the size of the excitation is of the order of the radial size of the BEC.

The simulations for small density perturbations ($\Delta n/n=10^{-4}$) confirm the minimum of the density dip moves with $c_\text{B}(0)$ as given by \Eq{bogolsostrap} and shows the validity of the simulation. For the perturbation used in the experiment the simulations show the minimum of the dip propagates about 8\% slower than the speed of sound, in agreement with the estimate  $\Delta c = -(c/2) \Delta n/n$. This is disadvantageous for the determination of the speed of sound from the measured density profiles, since the position of the minimum of the dip is the easiest parameter to derive. In order to derive the appropriate speed of sound from the density profiles, not only the position of minimum is determined, but also the amount of deformation is taken into account by determining the asymmetry of the propagating wave.

The simulations suggest the density at the ($1/\Eu$)-height of the dips is a good approximation of the unperturbed density (\cf \Fig{simdens}). We therefore use the deformation at $(1/\Eu)$-height in addition to the position of the center of the perturbation in these simulations as the measure of the distance the perturbation has traveled. The speed of sound is now found to remain constant within 2\% for a broad range of perturbation amplitudes ($10^{-5}$--$0.3$), only to show larger deviations when shock waves are formed. For typical propagation time used in the experiment, the simulations show the decrease of the propagation speed due to the axial density dependence remains below 2\% as well. 

Returning to the results of the experiment, the measured density profiles are fitted to an asymmetric function, yielding both the deformation and position of the propagating condensate density wave. Taking the same combined measure for the traveled distance as used in the simulations, the propagation distance varies linear with the propagation time, the slope of which is used to determine the speed of sound.

This procedure is repeated for different temperatures, which are created by adjusting the final rf-field frequency. Each series, consisting of about ten shots, is used to determine the propagation speed for that temperature. Rf-induced evaporative cooling does not allow to cool to temperatures below $\kb T \approx \mu$, since around this temperature both thermal atoms and condensed atoms are removed from the trap by the rf-field, setting a lower limit to the temperature reached in the experiment. The highest temperature used in the measurements corresponds to $T/T_\text{c} = 0.72 \pm 0.04$. For higher temperatures the axial length of the BEC becomes more than 15 \% shorter than the typical BEC length. By choosing this upper limit for the temperature prevents the need to account for the axial density dependence.

\begin{figure}[htbp]
\begin{center}
$\begin{array}{c}
  \includegraphics[width=0.45\textwidth]{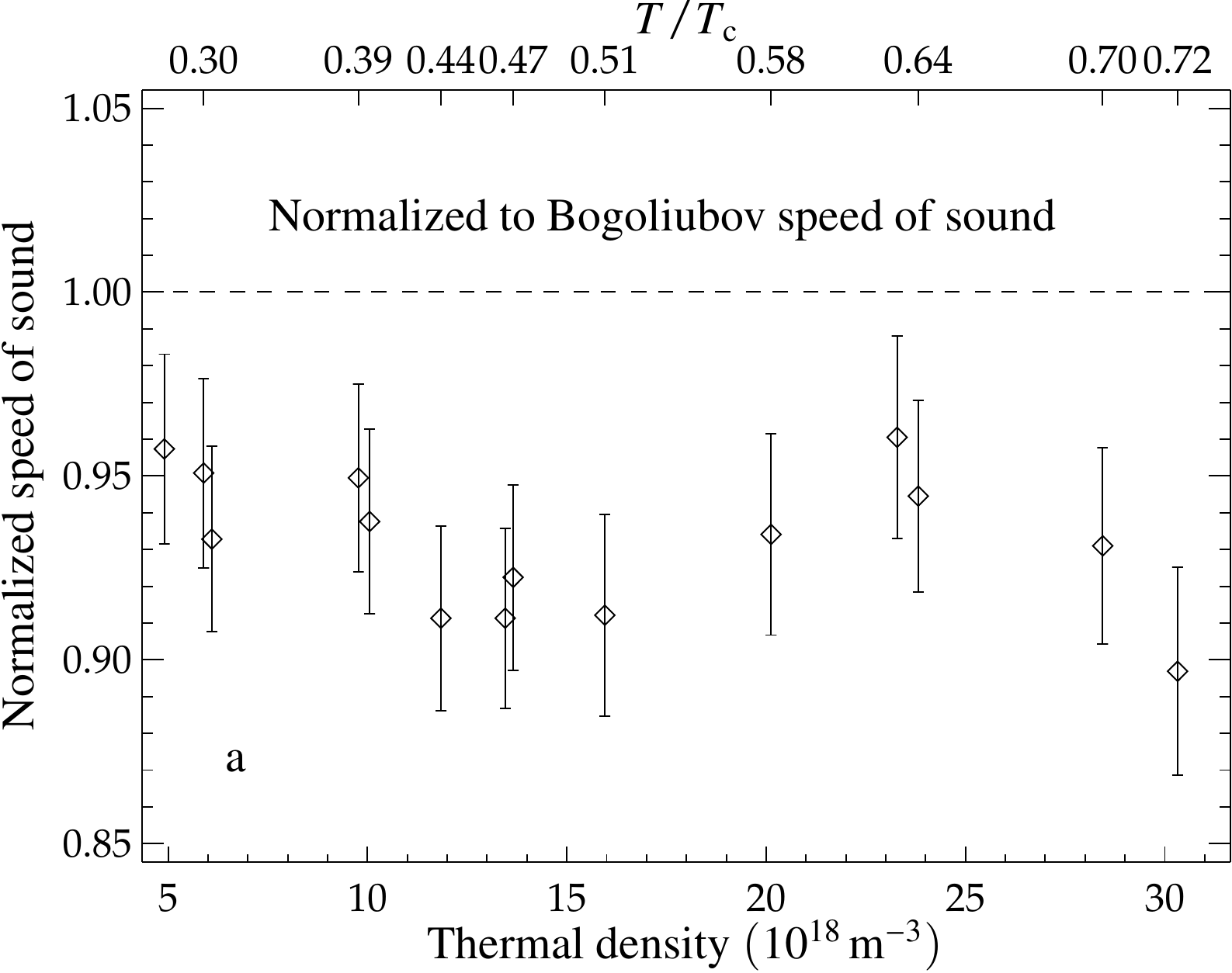}  \\ 
  \includegraphics[width=0.45\textwidth]{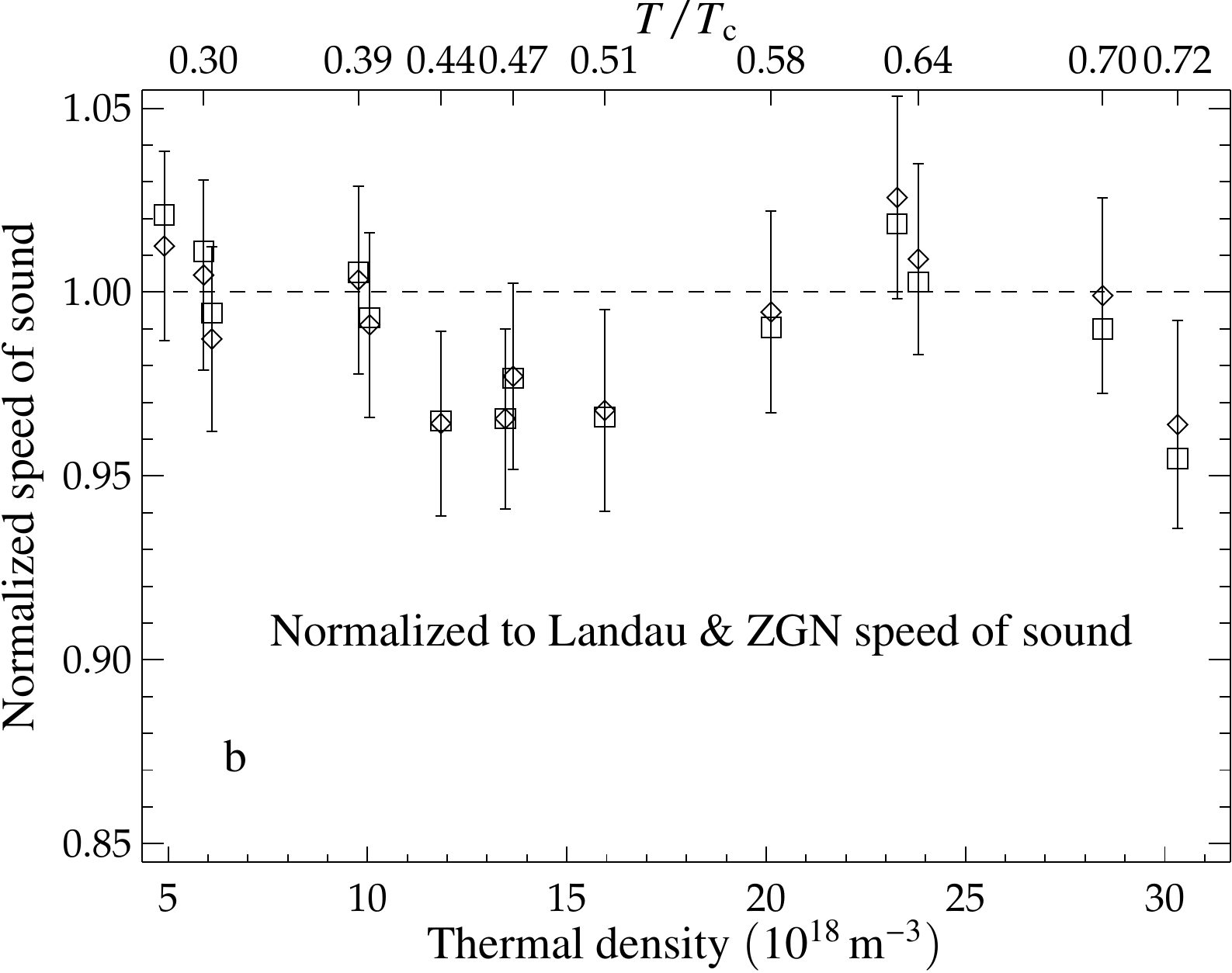} \\ 
\end{array}$
\end{center}
\caption{The normalized speed of sound as a function of the thermal density. The upper axis gives the reduced temperature $T/T_\text{c}$ for the corresponding data point. Figure (a) shows the speed of sound normalized to $c_\text{B}(0)$ (\Eq{bogolsostrap}) based on the measured BEC density. Figure (b) shows the same measurements normalized to the speed of sound based on the Landau model (diamonds) and ZGN model (squares) based on the measured BEC and thermal densities. The main contribution to the error bars is the uncertainty in the measured densities.}
\label{fig:sos-dens-temp}
\end{figure}

An unperturbed cloud in each series is used to derive the temperature, thermal density and BEC density for that series. In these measurements the chemical potential based on the axial size of the BEC and the total accumulated phase agree within 5\%. The measured speeds are normalized to $c_\text{B}(0)$ based on the measured BEC density. The resulting normalized speed of sound as a function of reduced temperature and thermal density is shown in \Fig{sos-dens-temp} (a). From these measurements we conclude that, even though the thermal density is varied over more than an order of magnitude, the effect of the thermal cloud appears to be constant within the accuracy of the measurement. Furthermore, the measured speeds are found to be approximately 7\% lower than $c_\text{B}(0)$ given by \Eq{bogolsostrap}.  Note that since the condensate density is limited by three-body losses, the variation of \nc as a function of the temperature is modest. 

In \Fig{sos-dens-temp} (b) the measured speeds are normalized using the speed of second sound calculated using the Landau and ZGN model. We find the measured propagation speed to be in excellent agreement with the speed of second sound given by both the Landau and ZGN model within the accuracy of the measurements. This result shows we have measured the effect of the thermal cloud on the propagation of a sound wave in the BEC. However, the dependence of the speed of second sound on the temperature is modest in the experimentally accessible temperature range. Furthermore, since the difference between the Landau and the ZGN model is smaller than the experimental uncertainty we cannot distinguish between both models in the current experiment. 

We have not been able to observe a sound wave in the thermal cloud (first sound) in this experiment. For the higher temperatures this can be explained by the moderate perturbation depth with respect to the thermal energy. At the lowest temperatures, still above $\kb T=\mu$, the thermal density is small and the thermal cloud spatially barely extends further than the BEC, causing the signal-to-noise to be insufficient to see small density perturbations in the thermal cloud in this regime. Above $T_\text{c}$ we are able to observe a density perturbation when the excitation is of the order $\kb T$, but the excited wave damps too fast to be able to observe its propagation.

\section{Conclusion}
In conclusion, we have excited a hydrodynamic mode in a BEC: a propagating sound wave. We measure its propagation speed, which is used to determine the speed of sound in the BEC as a function of the temperature. The combination of the phase-contrast imaging technique and elongated large atom number BECs allows us to make a moderate density excitation in the BEC and observe the propagation of a sound wave. Numerical simulations are conducted to model the non-linear propagation. We find the speed of sound to be in good agreement with both the Landau model and ZGN model in which the coupling between the first and second sound modes is incorporated and thus we have observed the effect of the thermal cloud on the speed of sound in the BEC. However, the effect does not vary within the accuracy of the measurements with the temperature for our experimental conditions. The Bogoliubov speed of sound, in which this coupling between the two sound modes is absent, deviates significantly from the measured speeds reported in this paper.

This work is supported by the Stichting voor Fundamenteel Onderzoek der Materie ``FOM'' and by the Nederlandse Organisatie voor Wetenschaplijk Onderzoek ``NWO''.

%\bibliographystyle{../../h-physrev3.bst}
%\bibliography{../../thesis}

\end{document}